\documentclass[10pt,a4paper]{article}

\usepackage[a4paper,left=20mm, right=20mm, top=30mm, bottom=30mm]{geometry}

\usepackage[utf8]{inputenc}
\usepackage{xspace}
\usepackage{amsmath}
\usepackage{amsfonts}
\usepackage{amssymb}
\usepackage{graphicx}
\usepackage{subcaption}
\usepackage{xspace}
\usepackage{url}
\usepackage{xcolor}
\usepackage{hyperref}
\usepackage{afterpage}
\usepackage{comment}
\usepackage{authblk}

\includecomment{qidoc}

\excludecomment{qidoc}

\usepackage{lineno}

\bibstyle{apacite}

\title{Quantum measurement fitting}
\author[1]{P.T. Eendebak}
\affil[1]{QuTech and Netherlands Organization for Applied Scientific Research (TNO), Stieltjesweg 1, 2628 CK Delft, Netherlands}

\date{\today}

\newcommand{\p}{\theta} 
\newcommand{\pp}{\ensuremath{p}\xspace} 
\newcommand{\x}{\ensuremath{x}\xspace} 
\newcommand{\y}{\ensuremath{y}\xspace} 
\newcommand{\ycount}{\ensuremath{k}\xspace} 
\newcommand{\fmtmodel}[1]{#1} 

\newcommand{\dof}{d} 
\newcommand{\np}{k} 
\newcommand{\shots}{N} 

\newcommand{\ypred}{\fit{y}}

\newcommand{\m}{m} 
\newcommand{\si}{s} 
\newcommand{\sn}{S} 

\newcommand{\OLS}{ordinary least-squares\xspace} 
\newcommand{\WLS}{weighted least-squares\xspace} 
\newcommand{\MLE}{maximum likelihood-estimation\xspace} 

\newcommand{\figureref}[1]{figure~\ref{#1}\xspace} 
\newcommand{\appendixref}[1]{appendix~\ref{#1}\xspace} 
\newcommand{\formularef}[1]{formula~\ref{#1}\xspace}

\newcommand{\rsubsection}[1]{{\bfseries #1}\xspace} 

\newcommand{\code}[1]{\texttt{#1}}

\newcommand{\fit}[1]{\tilde{#1}} 
\newcommand{\est}[1]{\tilde{#1}} 

\newcommand{\reg}[1]{{#1}_r} 

\newcommand{\estimate}[1]{\hat{#1}}

\DeclareMathOperator{\Penalty}{\mathrm{P}}
\DeclareMathOperator{\Jols}{\mathrm{J}_\mathrm{OLS}}
\DeclareMathOperator{\Jwls}{\mathrm{J}_\mathrm{WLS}}
\DeclareMathOperator{\Jmle}{\mathrm{J}_\mathrm{MLE}}
\DeclareMathOperator{\logr}{\reg{\log}}

\DeclareMathOperator{\MV}{N_\sigma}

\DeclareMathOperator*{\argmin}{arg\,min}
\DeclareMathOperator{\clip}{clip}

\usepackage{listings}
\usepackage{verbatim}

\iffalse
\usepackage{minted}
\newminted{python}{
    fontsize=\scriptsize,
    numbersep=8pt,
    frame=lines,
    framesep=3mm
}
\else
\newenvironment{pythoncode}{\verbatim}{\endverbatim}
\fi

\begin{document}
\maketitle

\begin{abstract}
Quantum measurements are not deterministic. For this reason quantum measurements are repeated for a number of shots on identically prepared systems.
The uncertainty in each measurement depends on the number of shots and
the expected outcome of the measurement. This information can be used to improve the fitting of models to quantum measurements.

In this paper we analyse ordinary-least squares, weighted least squares and maximum-likelihood estimation. We show that using the information on the quantum measurement uncertainty can lead to improved estimation of system parameters.
We also introduce the concept of model violation and demonstrate it can be a valuable
tool to analyze model assumptions and performance of quantum systems.

\begin{qidoc}
The methods described in this paper are available in the \code{qi-analysis} package.
See section~\ref{section:qlsi_implementation} for details.
\end{qidoc}
\end{abstract}

\section{Introduction}

Measurements on quantum systems often consist of a system initialization, manipulation and readout. This cycle is repeated multiple times to obtain a measurement histogram.
For calibrations of quantum systems often variations of the manipulation stage are measured, resulting in 1-dimensional datasets. An example is a Rabi oscillation which is used to calibrate the duration of an X90 gate~\cite[section 4.3]{Last2020}.

For a single-qubit quantum system in the state $\alpha |0\rangle + \beta |1\rangle$ the probability of finding the system in state $|1\rangle$ upon measurement is $p=|\beta|^2$. When measuring  identically prepared quantum states for $\shots$ times the number of counts $k$ of the $|1\rangle$ state is a value between 0 and $\shots$ (inclusive), distributed according to the binomial distribution which has probability mass function
\begin{align}
B(k, N, p) = \binom{\shots}{k}  p^k (1-p)^{N-k} .
\label{binom}
\end{align}
The mean and variance of this distribution are given by $\mu = p$ and $\sigma^2=p(1-p)/N$, respectively.

To obtain information about operational parameters of quantum systems we often perform calibration experiments where we fit models
with parameters to measured data points. Examples are a $T_1$ measurement (fitting an exponential decay), 
a $T_2^*$ measurement (fitting a damped sine wave) or a qubit spectroscopy measurement (fitting a Rabi model).

In this paper we compare several methods to fit the models to the data and analyse the properties
of the estimated parameters. In particular we pay attention to the bias and  variance of the estimated parameters.

\section{Methods}

In this work we consider a model $F(x, \p)$ that depends on a number of parameters $\p=(\p_0, \ldots, \p_k)$ and independent data variable $x$. An example is an exponential decay $F(x, \theta) = \p_o \exp( -\p_1 x)$ or a sine $F(x, \p) = \p_0 + \p_1 \sin(\p_2 x + \p_3)$.
In typical experiments we measure $\m$ datapoints $(\x_j, \y_j), 1 \leq j \leq m$ with $\y_j=\ycount_j/N$, with $\ycount_j$ the number of times the $|1\rangle$ state was measured.

\subsection{Fitting methods}

Most curve fitting methods depend on a cost function and a non-linear optimizer~\cite{Gavin2013TheLM}. In this section we focus on the cost function.
In ordinary least squares (OLS)~\cite{nonlinear_least_squares} one optimizes the cost function
\begin{align}
\Jols(\p) = \sum_{j=1}^\m | F(\x_j, \p) - \y_j | ^2 .
\label{OLS}
\end{align}
The OLS estimate for the unknown parameters is
\begin{align}
\estimate{\p}_\mathrm{OLS} = \argmin_\p J_\mathrm{OLS}(\p) .
\label{OLSestimate}
\end{align}
To find the estimate usually an initial guess is provided and a non-linear optimization method is applied to the cost function.

Alternative fitting methods are weighted least squares (WLS) and maximum-likelihood estimation (MLE).
The WLS cost function is 
\begin{align}
\Jwls(\p) = \sum_{j=1}^\m w_j | F(\x_j, \p) - \y_j | ^2
\label{WLS}
\end{align}
for a suitable choice of weights $w_j$. The choice of weights typically depends on the application. 
According to the Gauss-Markov theorem~\cite{GaussMarkovTheorem} using weights
$w_j = 1/\sigma_j^2$, with $\sigma_j^2$ the variance of the $j$th measured datapoint
results in the best linear unbiased estimator (BLUE estimator).

The MLE cost function is defined as the negative log-likelihood
\begin{align}
\begin{split}
\Jmle(\p) = - \mathcal{L}(\p) &= - \log \left( \prod_{j=1}^m B(k_j, N, F(x_j,\p) ) \right) \\
 	&=- \sum_{j=1}^\m  \log \left( \binom{N}{k_j} F(x_j, \p)^{k_j} (1-F(x_j, \p))^{N-k_j} \right) \\
 	&= \textrm{constant} - N \sum_{j=1}^\m  \left( \y_j \log( F(x_j, \p) ) + (1-y_j) \log( 1-F(\x_j, \p)) \right) ,
\end{split}
\label{MLE}
\end{align}
with $y_j$ the measured fraction for the $j$th datapoint.
The maximum-likelihood estimation optimizes the parameters to maximize the probability of observing the measured data.

The WLS method requires a choice of the weights $w_j$.
A natural choice is to take $w_j=1/\sigma_j$ with $\sigma^2_j$ the variance of the probability for the $j$th datapoint.
However, we do not know the real value of the probability a priori, so we need to make a suitable estimate. This process is called
bootstrapping~\cite{Wiki:bootstrapping}.
Both the WLS and the MLE methods require regularization. The regularization is required
to avoid the singularities in the cost functions for probabilities near 0 and 1. 
Another variation on the above methods is iteratively reweighted least squares (IRLS). In this method the WLS cost function is used, but weights are adapted on each iteration according to the current estimate of the parameters.

\subsubsection{Initial estimates and optimization}

The optimization of a cost function is performed by starting with an
initial guess of the parameters. In this paper we assume that we have a procedure available to produce a reasonable initial guess (either from the data or provided by the user). 
Starting with the initial guess the parameters of the model are optimized using a 
local optimization procedure such as gradient descent or Levenberg–Marquardt~\cite{LM}.

All results in this paper have been obtained with LMfit-py~\cite{lmfit} using the
\code{least-squares} method. We have not observed different results with other optimizers. 
The optimization with the non-OLS methods has been performed with initial parameters 
set to the output of the OLS fit.

\subsubsection{Bootstrapping}
\label{section:bootstrapping}

From the observed data $\y$ we want to obtain an estimate for the true probabilities $\pp_j$ and the variance
$\sigma_j^2$ for each datapoint.
To obtain an estimate for the values of $p_j$ and the variance $\sigma_j^2$  we have several options
\begin{description}
\item[Baseline] We take $\pp_j=\y_j$ and estimate the variance using the binomial $\sigma^2_j = \y_j(1-\y_j)/N$.
The variance is thus estimated from the measured datapoints.
\item[Jeffreys prior] We use the measured datapoints $y_j$ to update a prior distribution to a posterior distribution. A reasonable choice for the prior distribution is Jeffreys prior with equal probability for the 
$|0\rangle$ and $|1\rangle$ states (this choice for the prior is used in Qiskit~\cite{QiskitJeffrey}).
The observation of the datapoint $y_j$ then leads to estimates
\begin{align}
p_j &= \frac{N y_j + 1/2}{N+1},\quad \sigma_j^2= \frac{p_j(1-p_j)}{N+2} .
\label{bootstrap_qiskit}
\end{align}
\item[Wilson] From the estimates $\pp_j$ we determine the Wilson score interval~\cite{Wilson} for $z=1$.
We estimate the variance as the centre of the score interval.
\item[Prediction] We use the OLS estimates of the model parameters $\p$ to predict values $\fit{\y}_j$ and then
use $p_j=\y_j$, $\sigma^2_j= \fit{\y}_j(1-\fit{\y}_j)/N$. We can also use $p_j=\fit{\y}_j$ in combination with one of the methods described above. This creates more balanced estimates
for the variance and avoids using small variances for datapoints that are close to 0 or 1 due to
the random component in the measurements.
\end{description}
More variations are possible, but with these options we have introduced the main concepts used in bootstrapping.

\subsubsection{Regularization}

Some of the bootstrapping methods described above already introduce some regularization, but some others do not. Regularization is applied to either the probabilities $p_j$ or the variances $\sigma_j^2$ or both. The regularization has several goals:
\begin{itemize}
\item Prevent errors in the optimization at the singularities $p=0$ and $p=1$.
\item Prevent errors when a predicted probability is outside the $[0, 1]$ range. That can happen for certain models when the parameters are automatically varied by the optimizer. But also error mitigation techniques like readout error mitigation can introduce pseudo-probabilities with negative values.
\item Without regularization the optimizers tend to converge to local minima more often.
\end{itemize}
A good regularization should address the issues mentioned above, and at the same time
\begin{itemize}
\item Avoid bias in the fitting procedure
\item The regularization should be differentiable (which improves performance of local optimizers)
\end{itemize}
All regularization methods considered have a parameter $\epsilon$ that describes the regularization strength. Typically, the value is regularized for $p_j < \epsilon$, while values $p_j > \epsilon$ are hardly affected by the regularization (and similar for $p_j$ near 1).
We use the following regularized probability on our experiments:
\begin{align}
r(p, \epsilon) &= 
\begin{cases}
    \epsilon/2 & \quad \text{if $p<0$} \\
    \epsilon/2 + p^2/(2\epsilon) & \quad \text{if $0 \leq p \leq \epsilon$} \\
    p & \quad \text{if $\epsilon \leq p \leq 1-\epsilon $}  \\
    1- (\epsilon/2 + (1-p)^2/(2\epsilon)) & \quad \text{if $1-\epsilon \leq p \leq 1$} \\
    1-\epsilon/2 & \quad \text{if $p>1$} 
  \end{cases}  .
\label{regularized probability}
\end{align}
For the MLE estimation we also have a regularized log function $\reg{\log}$. 
The regularized log is a second order Taylor expansion at $\epsilon$ for small
values~\cite[section 4.3]{Nielsen2021gst}.
\begin{align}
\logr(x, \epsilon) &= 
\begin{cases}
    \log(\epsilon) + (x-\epsilon)/\epsilon - (x-\epsilon)^2 / (2\epsilon^2) & \quad \text{if $x<\epsilon$} \\
    \log(x) & \quad \text{if $x > \epsilon$} \\
  \end{cases}  .
\label{regularized log}
\end{align}
In contrast to~\cite[section 4.3]{Nielsen2021gst} we have models that can predict datapoints outside the $[0, 1]$ interval.
For MLE the regularization with the log probability is not enough to avoid predictions outside [0,1]. The reason is that the 
binomial probability density function for $y \approx 1$ is zero almost everywhere, rises sharply to a maximum at $p=1$ and is zero
for $p>1$. The regularization results in a term
\begin{align}
y \log_r(p) + (1-y)\log_r(1-p) \approx \log_r(p) .
\end{align}
The likelihood is maximal in the interval $[0, 1]$ for $p=1$. But values $p>1$ are rewarded,
so we need a penalty term. The penalty term should be high and sharp at the $p=0$ and $p=1$ boundary. An alternative is using $\log(r(p))$. This does not result in a strong penalty for $p$ outside $[0, 1]$, but at least the likelihood for $p \gg 1$ is not rewarded.
In this paper we use the soft penalty term
\begin{align}
\Penalty_\mathrm{soft}(p) = (\max(p, 1)-1 + \min(p, 0))^2 / \epsilon^3 .
\label{mle:soft}
\end{align}
An alternative is a hard penalty
\begin{align}
\Penalty_\mathrm{hard}(p) &= 
\begin{cases}
    0 & \quad \text{if $p \in [0, 1]$} \\
    L & \quad \text{otherwise} 
\end{cases} ,
\label{mle:hard}
\end{align}
for $L$ a large number, e.g. $L=1e100$.
The penalty for the \MLE fitting is the sum of the penalties for all the probability predictions.
In appendix~\ref{section:exponential_model} we show results of MLE fitting without a penalty,
the soft penalty from equation~\ref{mle:soft} and the hard penalty from equation~\ref{mle:hard}.

Another view of regularization is that it adds prior information~\cite{Dablander} to our problem.

\subsection{Evaluation criteria}

In section~\ref{section:results} we will analyze the performance of the methods described above.
We will do this by applying the fitting methods to simulated data.
For every simulation we perform, the fitted parameters should ideally be close to the parameters used in the simulation. However, due to the randomness in the generation of the data points, it is not possible for the fitting to always match the ground truth parameters.
For every simulation  $s$ we obtain an estimate of the parameters
$\estimate{\p}_s$. For every method we consider the following properties
of the estimator:
\begin{align}
\mathrm{bias} &= (\sum_\si \estimate{\p}_\si - \p) / \sn , \\
\mathrm{variance} &= \sum_\si |\estimate{\p}_\si - \p|^2 / \sn , \\
\mathrm{likelihood} &= (1/\sn) \sum_\si \log \mathcal{L}(\estimate{\p}_\si) , 
\end{align}
with the summation $\si$ over the different simulations and $\sn$ the the number of simulations performed.

From our fitting procedure we want to have a small bias, low variance and high likelihood. These three properties cannot always be optimized simulatenously.
In particular there can be a trade-off~\cite{wikipedia_bias_varianc_tradeoff} between 
the bias and the variance of an estimator.
In this work we present results on the bias and variance of the estimators. The likelihood  of the estimate, and in particular whether the estimate is the MLE estimate, is considered less important.

\subsection{Model violation}

A measure of the deviation of the measured data to the model predictions is the $\chi^2$ value.
When normalized using the number of degrees of freedom $\dof = \m-\np$, we have the model violation $N_\sigma$.
\begin{align}
\chi^2 &= \sum_j \frac{(\y_j-\ypred_j)^2}{\sigma_j^2} = N\sum_{j=1}^\m \frac{(\y_j-\ypred_j)^2}{\ypred_j(1-\ypred_j)} \\
N_\sigma &= \frac{\chi^2-\dof}{\sqrt{2\dof}}
\label{chi2}
\end{align}
The model violation in the context of gate set tomography is described
 in~\cite[section 4.3]{Nielsen2021gst}.
%
%
%
For a sufficient amount of data points $\m$ the $\chi^2$ value is distributed
as a $\chi^2$-distribution with $\dof$ degrees of freedom. The model violation $N_\sigma$
is then normally distributed and large values of $\MV$ indicate the measured data cannot be explained by the model and fitted parameters.
Any outliers in the data will result in a high model violation. There are techniques
to make the fitting robust against the outliers
 (e.g. robust cost functions~\cite[section A6.8]{Hartley2003}, RANSAC~\cite{RANSAC}), but they are not in the scope of this work.

\section{Results}
\label{section:results}

In this section we compare the different fitting methods. We analyse OLS, WLS with different choices for the weights, and MLE. To obtain weights for the WLS we use the bootstrapping methods described in section~\ref{section:bootstrapping}.

We generate results by selecting a model $F$, simulation parameters $\p$, independent data points $x$ and a number of shots $\shots$. For a large number of iterations we generate datapoints according to our model, perform fitting of the parameters using one of the methods describing above and compare the fitted parameters $\fit{\p}$ with the simulation parameters $\p$.

\begin{figure}[ht]
  \centering
    \includegraphics[width=0.7\textwidth]{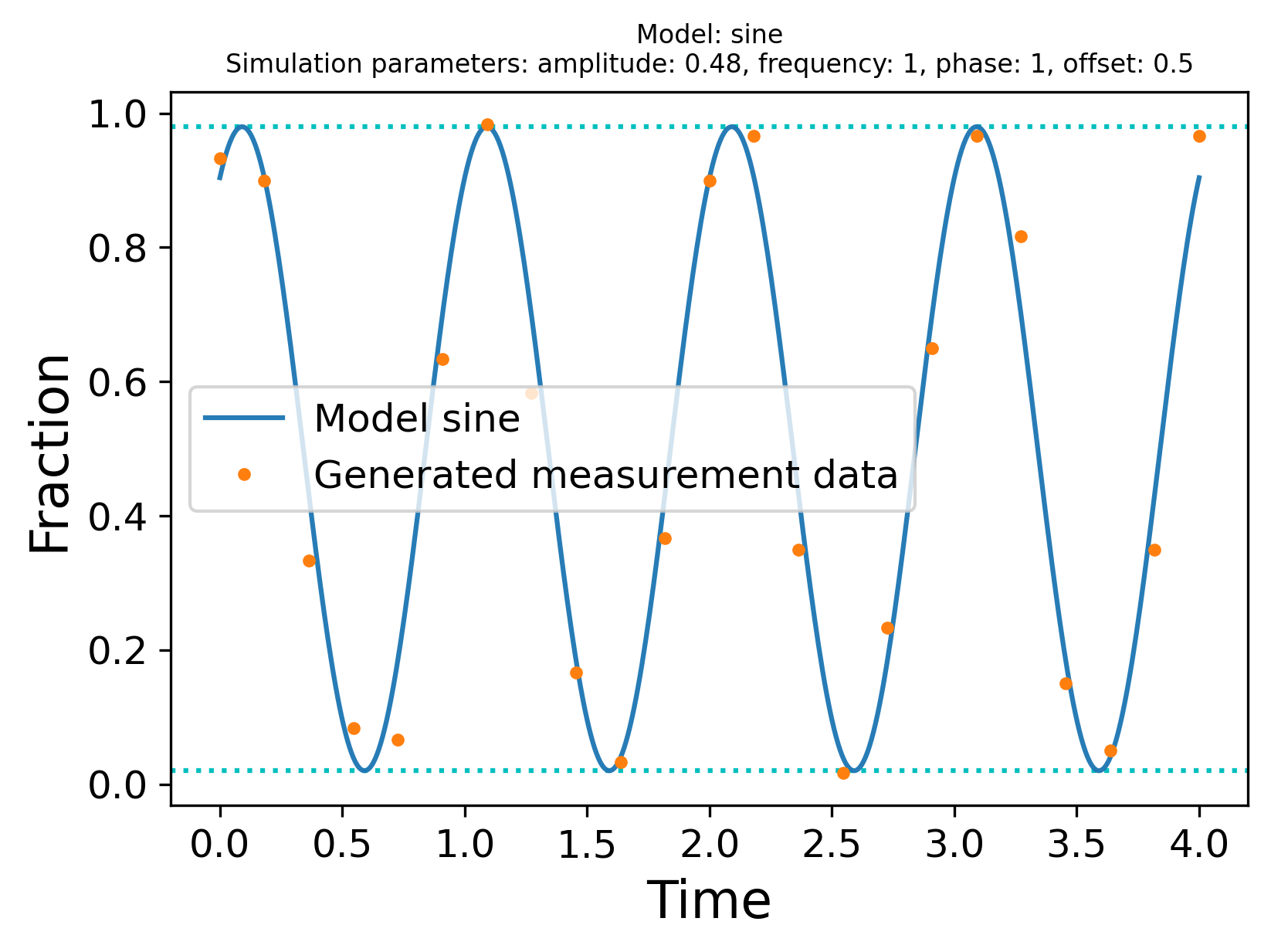}
  \caption{Model and typical measurement data for \code{sine}.}
  \label{figure:fitted_parameter_distribution_sine}
\end{figure}

The main example we use is a \fmtmodel{sine} model
\begin{align}
F(x, \p) = A \sin(2\pi f x + \phi) + \textrm{offset} ,
\end{align}
with parameters $\p=(A, f, \phi, \textrm{offset}) =  (0.48, 1, 1, .5)$.
The independent datapoints $x$ have been chosen to be 23 linearly spaced datapoints in the $[0, 4]$ interval.
The model with a set of simulated datapoints is shown in figure~\ref{figure:fitted_parameter_distribution_sine}.

The number of shots we focus on is either a small number of shots $N \sim 60$ or a moderate number of shots $\shots \sim 1000$ (in quantum experiments on hardware 100 to a 1000 shots  is a typical number). 
For either a high number of datapoints or a high number of shots
asymptotic results guarantee that all the estimators are unbiased, have minimum variance and are asymptotically normal~\cite{mle_normality}.
The regularizations are performed with parameter $\epsilon = 0.05 / \shots$, with $\shots$ the number of shots from the experiment.

\begin{figure}
  \centering
    \includegraphics[width=0.92\textwidth]{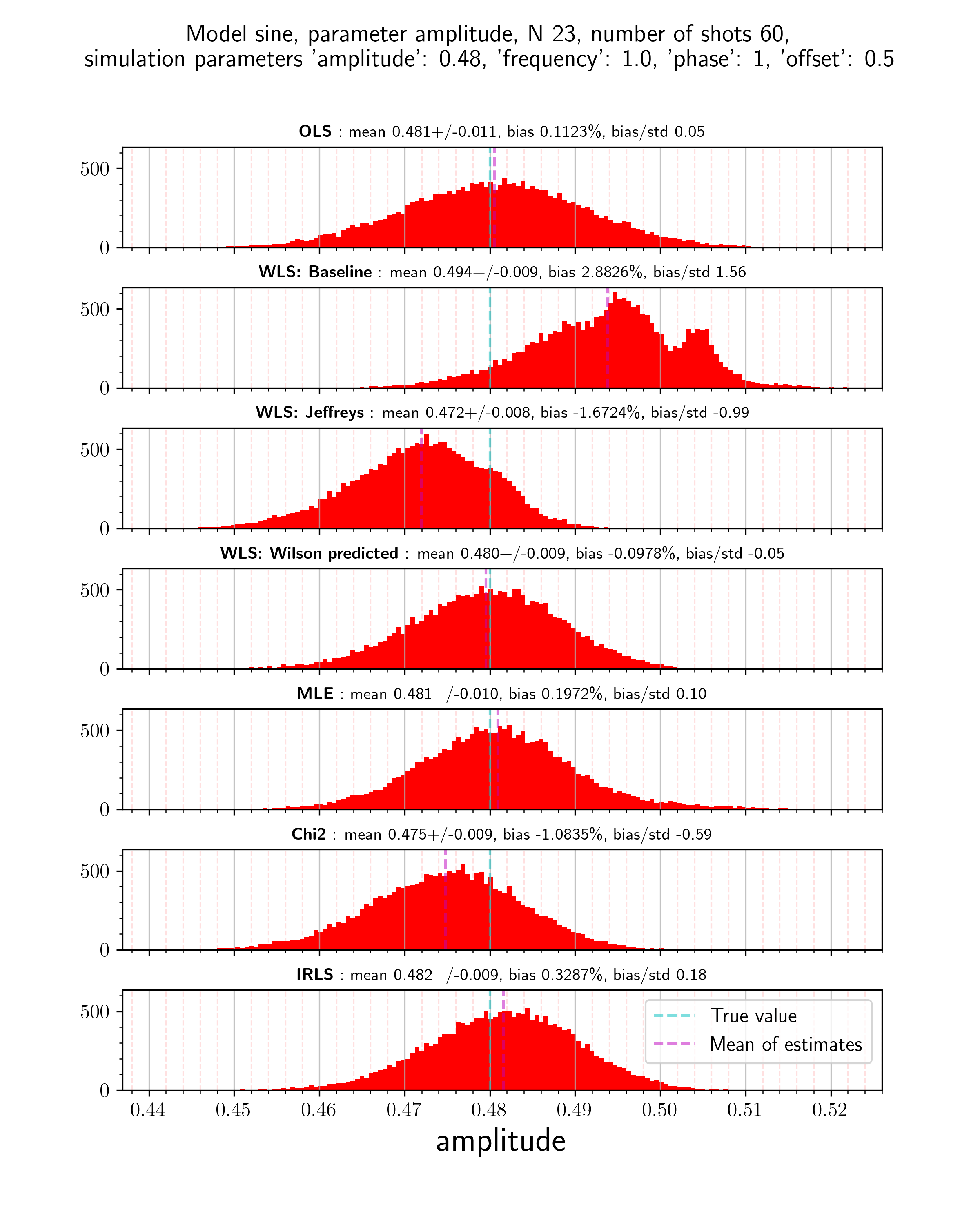}
  \caption{Distribution of the \code{amplitude} parameter for a \code{sine} model. For each method
   we show the mean and standard deviation of estimated parameter $\est{\p}$,
  the bias with respect to the simulation parameter and
  the ratio of the bias and the variation.}
  \label{figure:fitted_parameter_distribution}
\end{figure}

\rsubsection{Bias and variance}
In figure~\ref{figure:fitted_parameter_distribution} we show the distribution of the
estimated \code{amplitude} parameter for the \fmtmodel{sine} model.
The \OLS method has an estimator with small bias and a variation
of about 0.5\% in the amplitude.

The weighted least-squares based on the baseline bootstrapping performs poorly. The reason is that datapoints with probability near 0 and 1 can generate measured values very close to 0 or 1 leading to extreme values of the weights. We therefore recommend not to use this approach.

\begin{figure}
  \centering
    \includegraphics[width=0.92\textwidth]{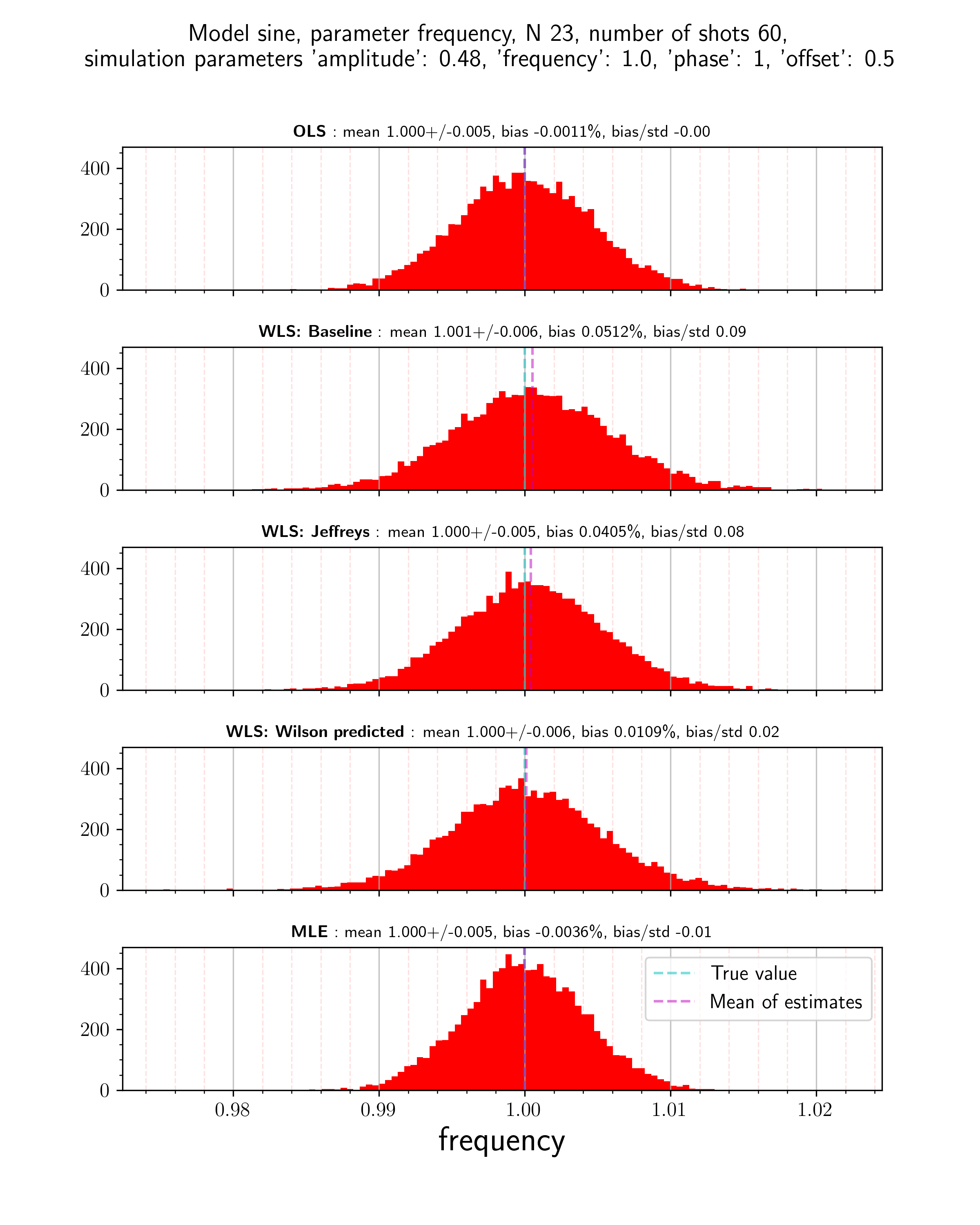}
  \caption{Distribution of the \code{frequency} parameter for a \code{sine} model. For each method
 we show the mean and standard deviation of estimated parameter $\est{\p}$,
   the bias with respect to the simulation parameter and
  the ratio of the bias and the variation.}
  \label{figure:sin_frequency_distribution}
\end{figure}

The WLS method with Jeffreys priors has a bias in the amplitude. The reason for this bias is that instead of fitting to the measurement datapoints $y_j$,
we fit to datapoints $(y_j + 0.5)/(\shots+1)$ (\formularef{bootstrap_qiskit}) which are always closer to 0.5 than the original datapoints $y_j$.
In the simulation a low number of shots $\shots=60$ has been used, leading
to a reduction of the amplitude of datapoints by a factor of $1.6\%$. This is reflected in a bias of 
the estimated amplitude. 

Both the WLS with predicted variance and the \MLE perform better than the \OLS, but the improvement is only modest. The variance is smaller
and there is a smaller tail in the distribution.
The $\chi^2$ based fitter (using $\chi^2$ as a proxy for the log-likelihood) is biased. This was already noted in~\cite[Appendix E]{Nielsen2000},
in the context of gate set tomography.
The iteratively reweighted least-squares (IRLS) has a bias towards a lower amplitude.
The reason for this is that higher amplitudes result in higher weights and therefore in general higher values of the cost function.
This favors the method to avoid high amplitudes.

In a similar way we find the estimate of the frequency parameter is an unbiased estimate
for all the methods, with standard deviation of about 0.5\% (\figureref{figure:sin_frequency_distribution}). To put this number  into perspective: to obtain
a single-qubit fidelity on an $X$ gate of 99.9\% one needs an accuracy in the frequency estimate of 2\%~\cite[Table I]{vanDijk2019}.

\begin{figure}
  \centering
 \begin{subfigure}{0.47\textwidth}
  \centering
    \includegraphics[width=0.87\textwidth]{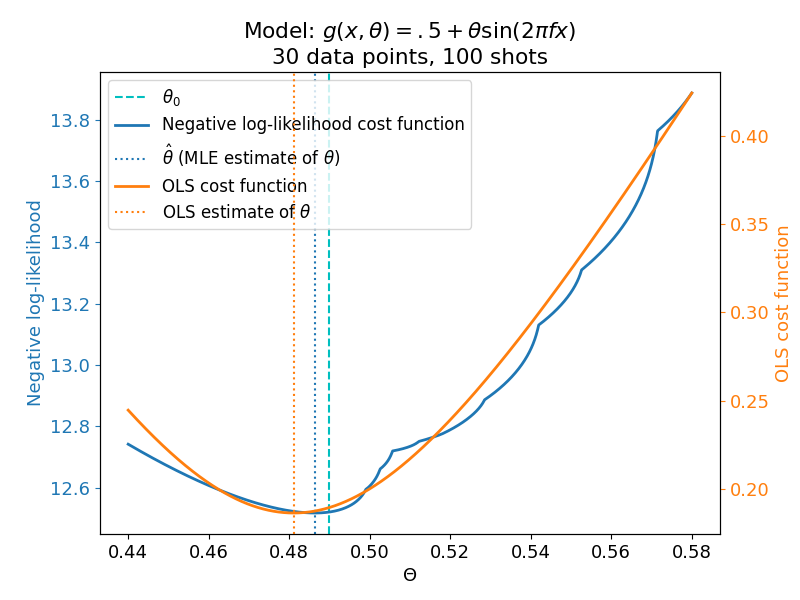}
    \caption{}
    \end{subfigure}
 \begin{subfigure}{0.47\textwidth}
  \centering
    \includegraphics[width=0.87\textwidth]{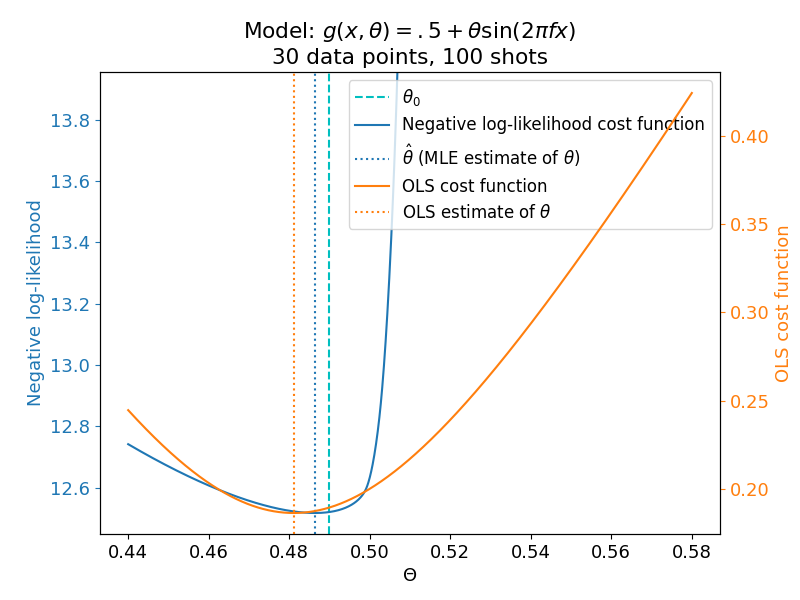}
    \caption{}
    \end{subfigure}
  \caption{Cost function of a sine model with the OLS cost function and a MLE cost function (negative regularized log-likelihood). The regularization in a) is with a regularized probability $\reg{p} = \clip(p, \epsilon, 1-\epsilon)$ and in 
  b) with the regularized log function $\logr$ (equation~\ref{regularized log}).}  
  \label{cost_function_regularization}
\end{figure}

\rsubsection{Importance of regularization and bootstrapping}
Without suitable regularization the cost functions can have local minima or flat plateaus, leading to  non-optimal solutions or slow convergence. An example of the regularization
with a non-differentiable method is shown in figure~\ref{cost_function_regularization}. The effect of the regularization for the \code{exponential} model is analysed in further detail in \appendixref{section:mle_penalty}.

\begin{figure}
	\centering
	\includegraphics[width=0.7\linewidth]{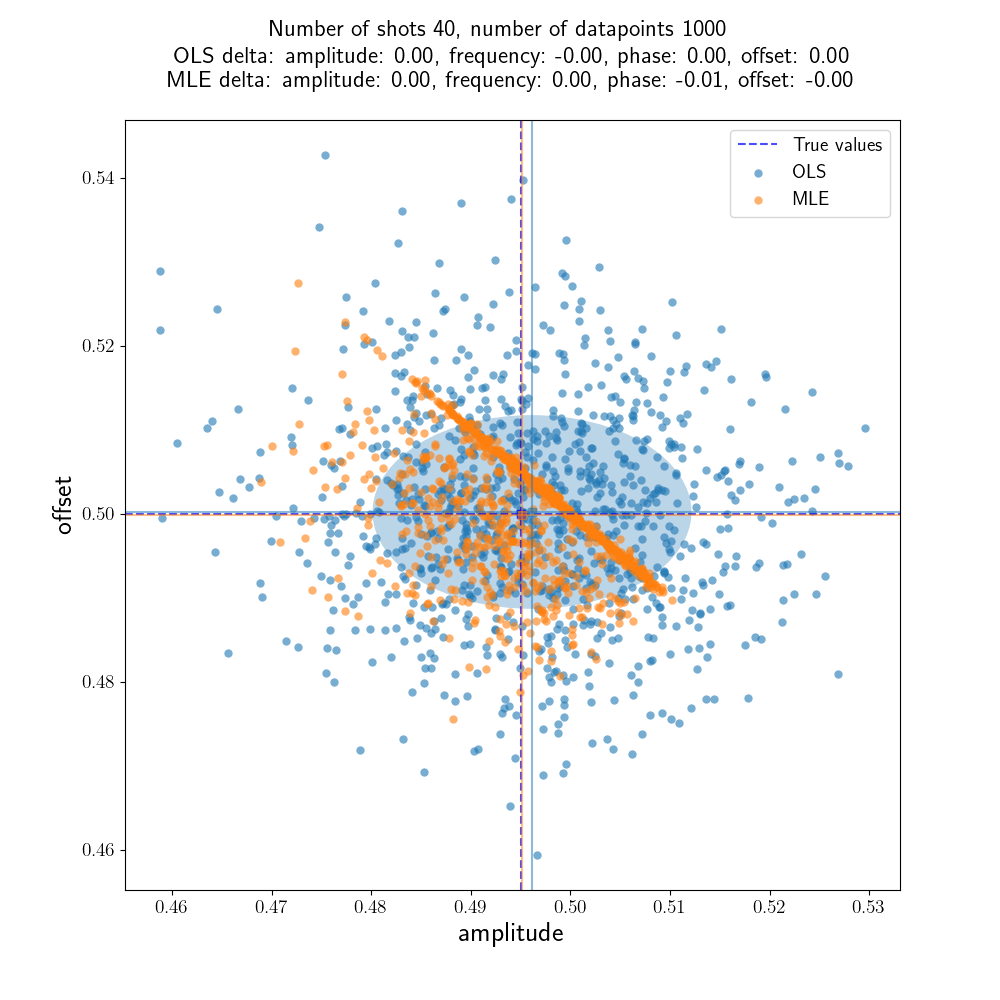}
	\caption{Scatter plot of OLS and MLE fits. The MLE has a penalty cost for data points outside the region of datapoints in the probability region $[0, 1]$. The result is a cluster of datapoints along the line \code{offset+amplitude=1}.}
	\label{fig:correlation_amplitude_offset_OLS_MLE}
\end{figure}

\rsubsection{Correlations between parameter estimates}
When making a scatter plot between two of the parameters we can see correlations between the parameters.

For example in~\figureref{fig:correlation_amplitude_offset_OLS_MLE} one can see a sharp line of MLE estimates with hardly any estimates above.
The reason for this is the following: if the sum of the \code{offset} and \code{amplitude} is larger than one, our model predicts
a measured fraction outside the valid probability range [0, 1]. The OLS method allows these terms (with the usual quadratic cost),
but for the MLE estimator this results in a strong penalty term (determined by the regularization).

\rsubsection{Number of samples and data points}
The variance in the estimated parameters depends on the number of shots and the number of datapoints. More samples or more data points lead to better estimates. However, for practical reasons it is desirable to keep both small.

\begin{figure}
  \centering
  
    \includegraphics[width=0.99\textwidth]{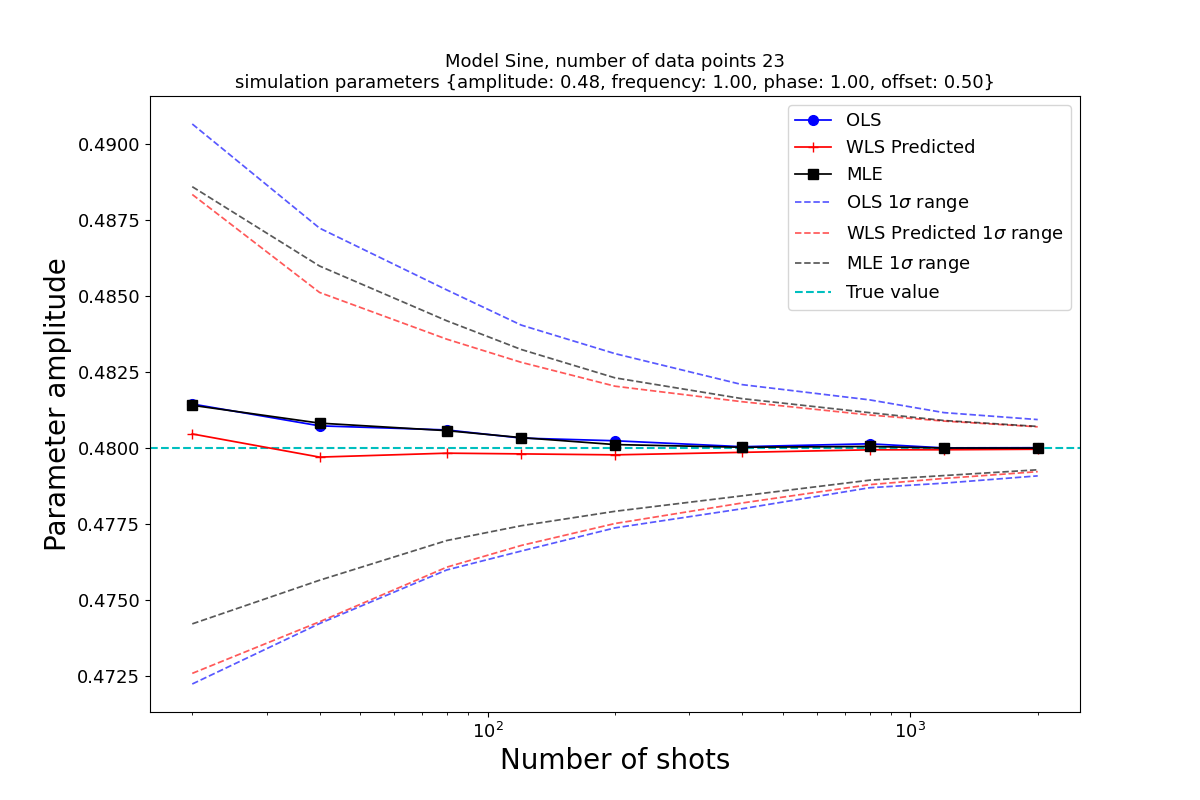}
  
  \caption{Dependence of the fitting accuracy on the number of datapoints. For different fitting methods
  	we have simulated 8000 experiments and calculated the mean and standard deviation of the fitted parameters.
  	 As expected the accuracy roughly scales as one over the square root of the number of shots.}  
  \label{fitting_N_variation}
\end{figure}

In \figureref{fitting_N_variation} we show
how the mean and variance of the estimate depend on the number of shots.
For a large number of shots the three different methods
have estimators with the same mean and similar variance.
The WLS method performs best overall in terms of the variance.

\begin{figure}
  \centering
 \begin{subfigure}{0.82\textwidth}
  \centering
    \includegraphics[width=0.7\textwidth]{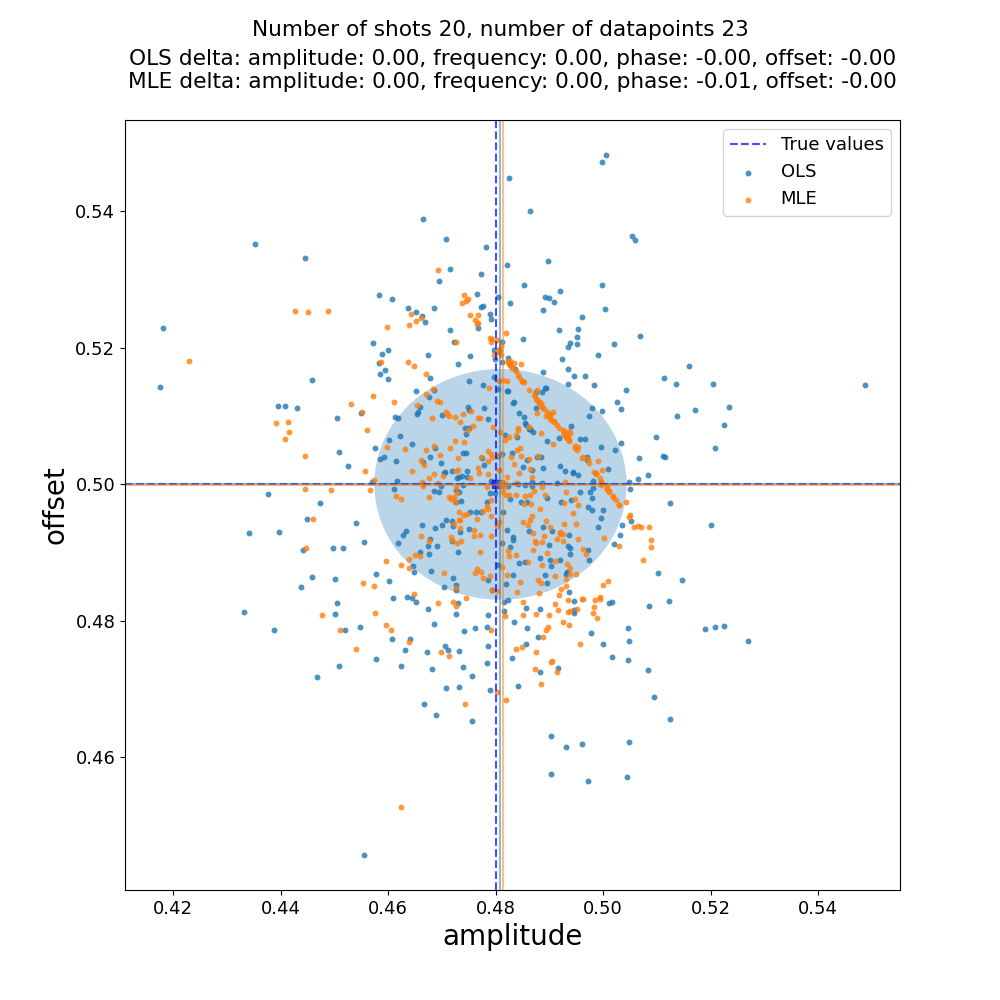}
    \caption{Small number of shots and small number of samples. The MLE fitting shows fewer outliers with respect to the value of the true parameters. Also the boundary of the valid probability predictions is visible.}
    \end{subfigure} 
 \begin{subfigure}{0.82\textwidth}
  \centering
    \includegraphics[width=0.7\textwidth]{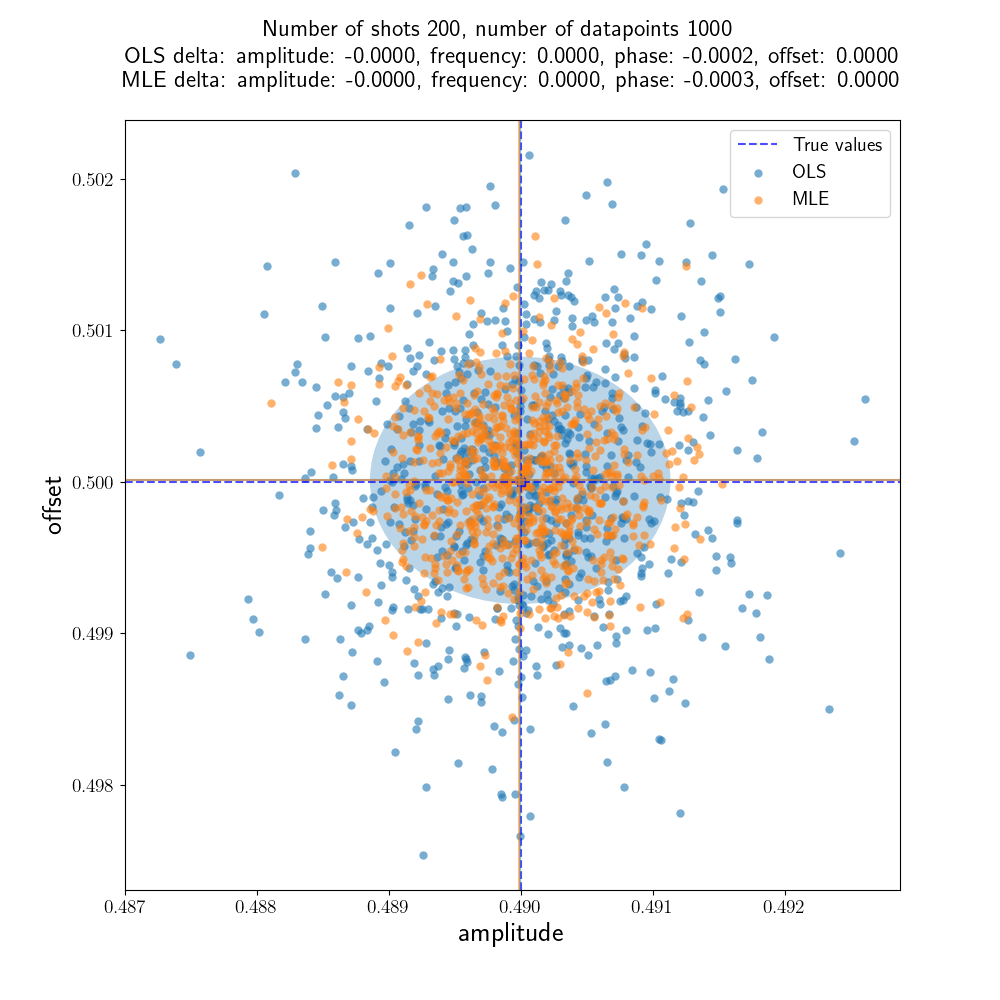}
    \caption{Large number of samples. The MLE estimates show no outliers, and have a smaller variance than the OLS estimates.}
    \end{subfigure}
  \caption{Fitting a sine model with OLS (red) and MLE (blue). The fitted \code{amplitude} and \code{offset} are shown in a scatter plot.}  
  \label{MLElimit}
\end{figure}

The MLE estimation is efficient in the limit of a large number of samples. This is shown
in figure~\ref{MLElimit}. For typical quantum experiments the number of data points is less than a hundred,
so we are in the region where we transition from outliers (due to a small number of samples)
and an efficient estimator (in the large number of samples limit).

\begin{figure}
  \centering
    \includegraphics[width=0.72\textwidth]{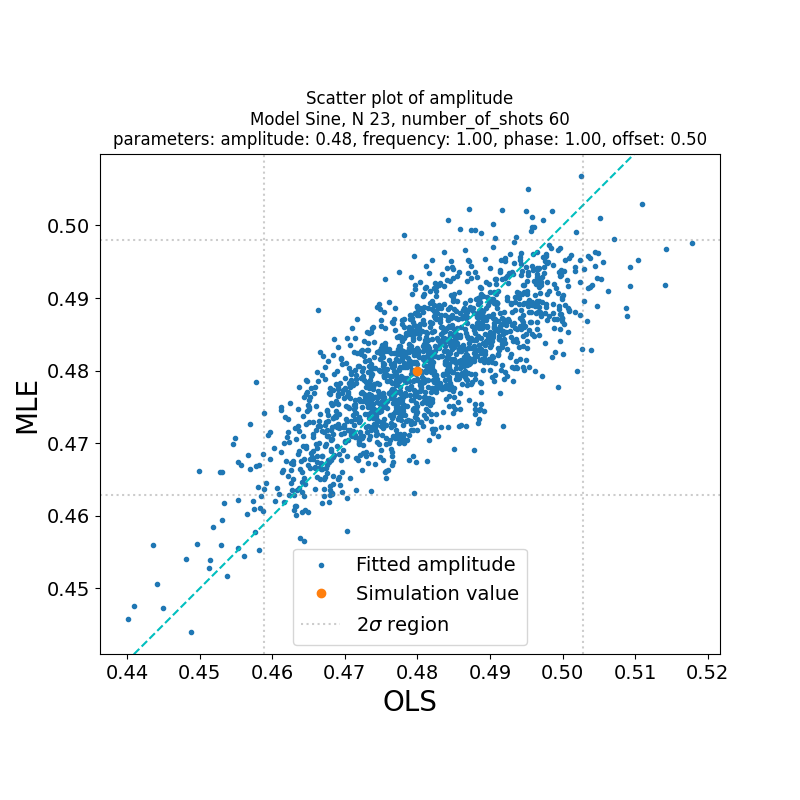}
  \caption{Scatter plot of the \code{amplitude} parameter for OLS and MLE. The difference between the OLS and MLE estimate
  	 of \code{amplitude} has zero mean and standard deviation of 0.0068.}
  \label{figure:scatterOLSMLE}
\end{figure}

\begin{figure}
  \centering
    \includegraphics[width=0.77\textwidth]{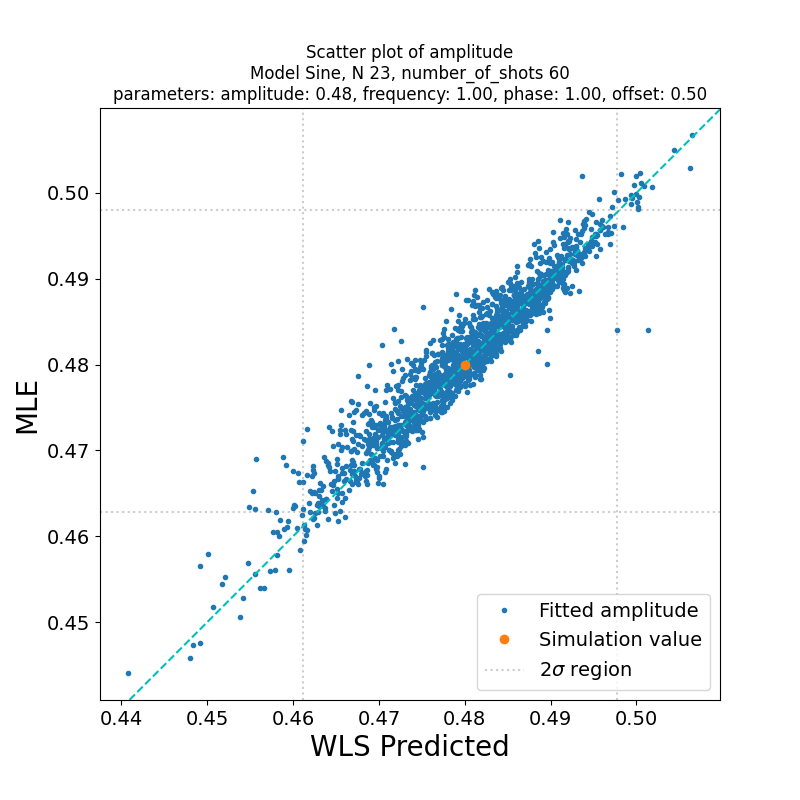}
  \caption{The amplitude is sensitive to regularization of the data points or the variance. Most of the time there is a very strong correlation between the results.
  But there are cases where there is a large difference due to different local minima}
  \label{figure:scatterMLEWLS}
\end{figure}

\rsubsection{Correlation between estimates of different methods}
The values of the fitted parameters with respect to the true parameter values
depend on the data points. There is a strong correlation between
the different methods (figures \ref{figure:scatterOLSMLE}
and~\ref{figure:scatterMLEWLS}).
For a given model the MLE estimates of a parameter can be both higher and lower than the OLS estimates. But in general a high estimate for OLS implies also a high estimate for WLS or MLE.

\rsubsection{Model violation as a diagnostic tool}
%
The model violation $N_\sigma$ can be used as a tool to identify problems with quantum systems or model assumptions. We give two examples from actual quantum measurements.

\begin{figure}[ht]
  \centering
 \begin{subfigure}{0.465\textwidth}
  \centering
    \includegraphics[width=\textwidth]{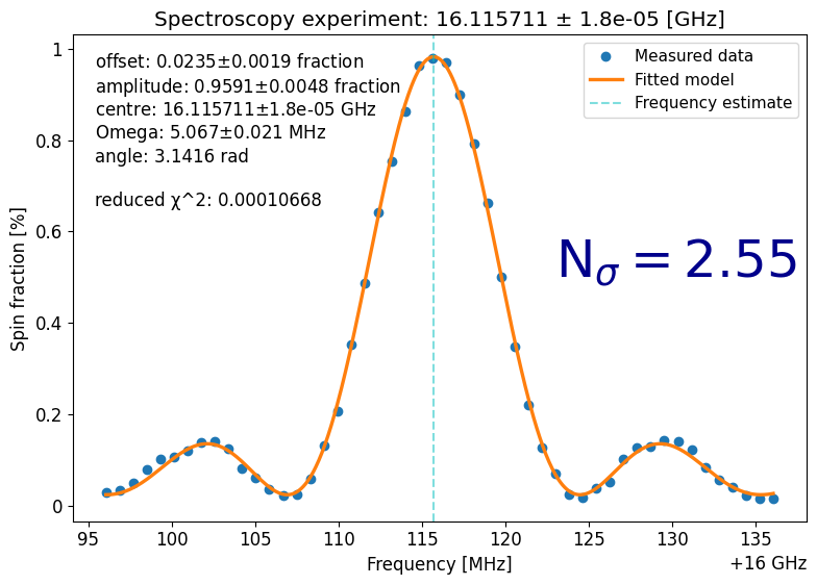}
    \caption{}
    \end{subfigure}
 \begin{subfigure}{0.465\textwidth}
  \centering
    \includegraphics[width=\textwidth]{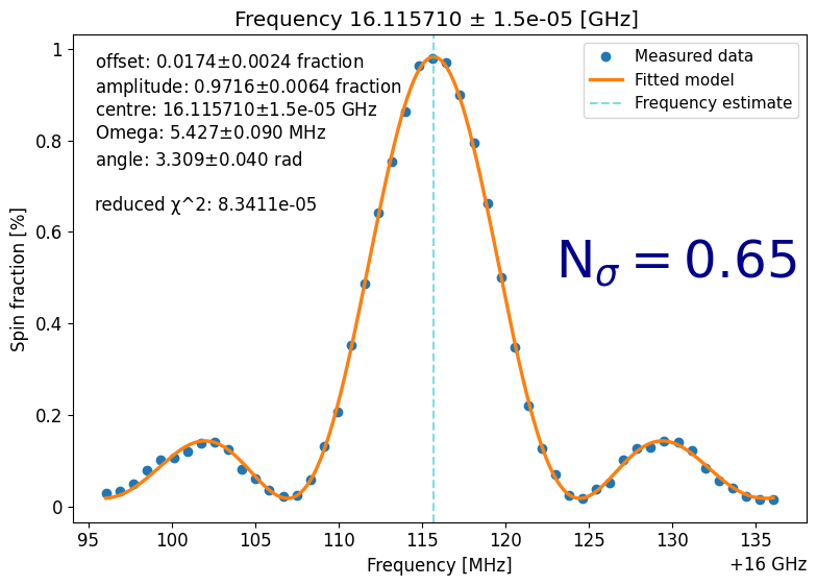}
    \caption{}
    \end{subfigure} 
  \caption{Model violation for a spectroscopy experiment. In (b) the rotation angle $t$ is estimated to be $1.05\pi$ radians.}
  \label{rabi_model_violation}
\end{figure}

The first example is data from the calibration sequence of a 6-qubit device~\cite[Figure 4c]{Philips2022} where a fit is performed in a spectroscopy experiment.
The model fitted is
\begin{align*}
F(x, \p) = A + B\frac{\Omega^2}{\Omega^2+(x-\omega_0)^2} \sin^2( \sqrt{\Omega^2+(x-\omega_0)^2} t/2 ) ,
\end{align*}
 with parameters $\p=(A, B, \Omega, \omega_0 )$ and $t=\pi$ the rotation angle.
The parameters are the offset $A$, visibility $B$, Rabi frequency $\Omega$ and 
the applied microwave frequency $\omega_0$.
In the spectroscopy experiment from~\cite{Philips2022} a standard fit 
with $t=\pi$ is performed. The measured data with the fitted model is shown
in figure~\ref{rabi_model_violation}.
The fitted data looks good at first sight, 
but the model violation (for $N=1000$)  $N_\sigma=2.55$ is on the high side. In particular the data and model seem to deviate at the side-lobes.
In this case it turns out that the experiment was performed with a small over-rotation of the pulse. If we repeat the model fitting, but also allow the angle parameter $t$ to vary, we find a model violation of $\MV=0.65$ and a rotation angle of $1.05\pi$.

\begin{figure}[ht]
	\centering
	\includegraphics[width=0.7\textwidth]{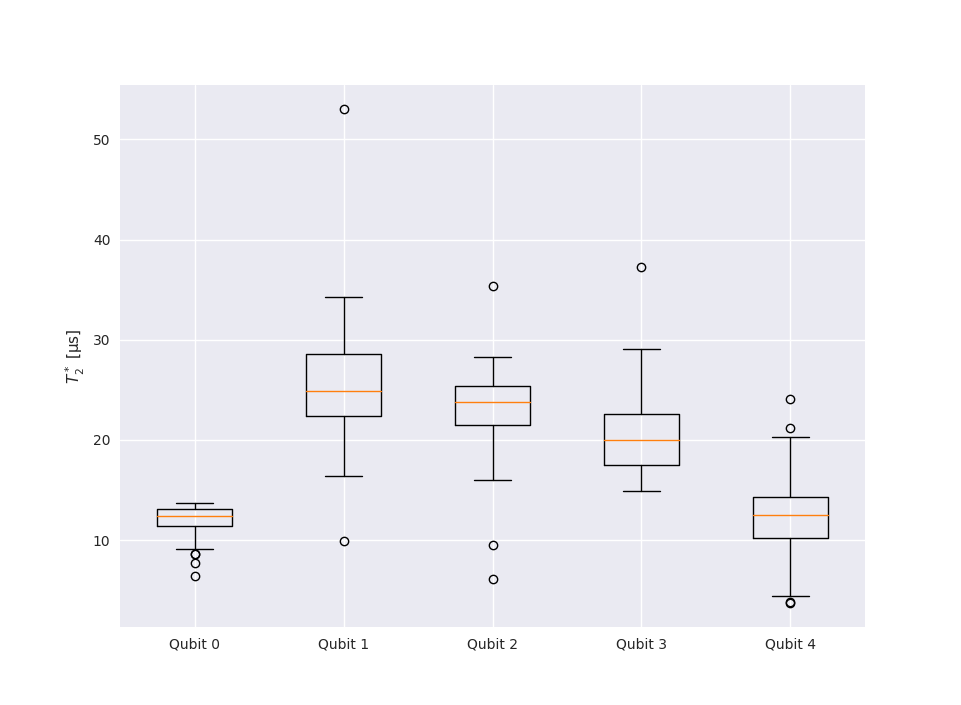}
	\caption{$T_2^*$ values for measurements on Starmon-5}
	\label{figure:t2star_boxplot}
\end{figure}

\begin{figure}[ht]
  \centering
    \includegraphics[width=0.7\textwidth]{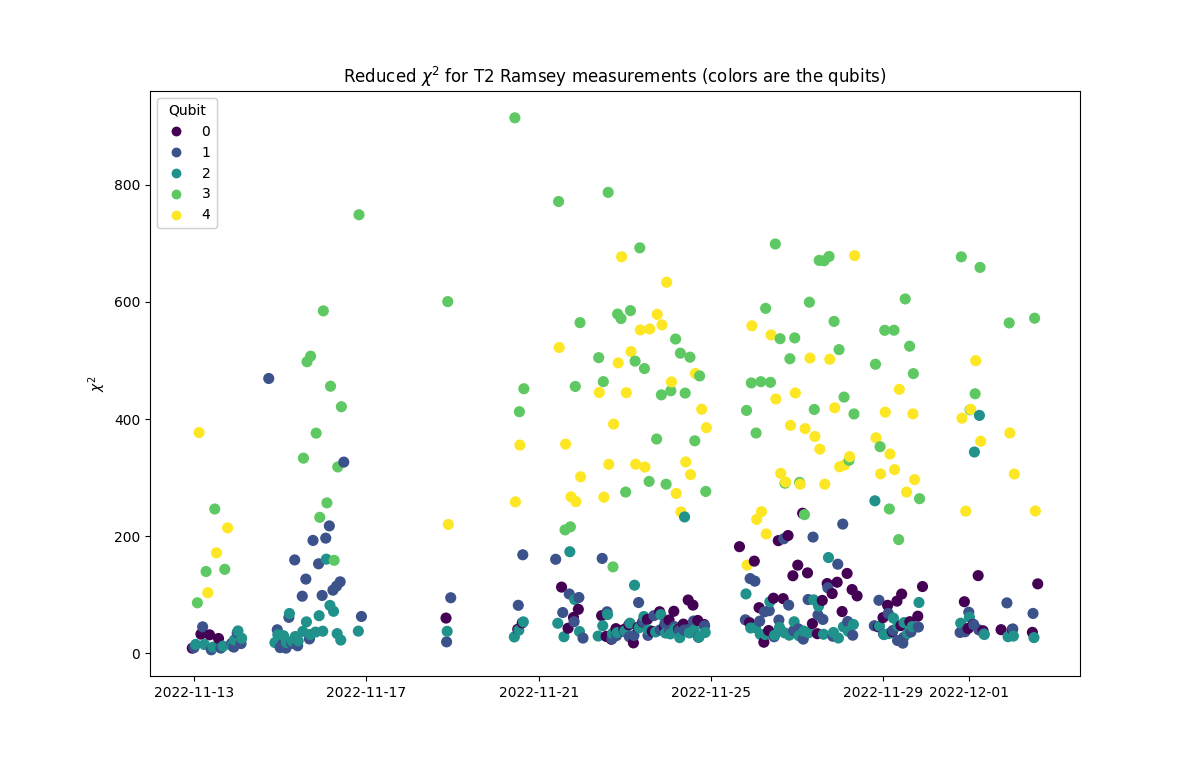}
  \caption{$\chi^2$ values for $T_2^*$ measurements on Starmon-5}
  \label{figure:chi2_vs_t2star}
\end{figure}

\begin{figure}
  \centering
    \includegraphics[width=0.7\textwidth]{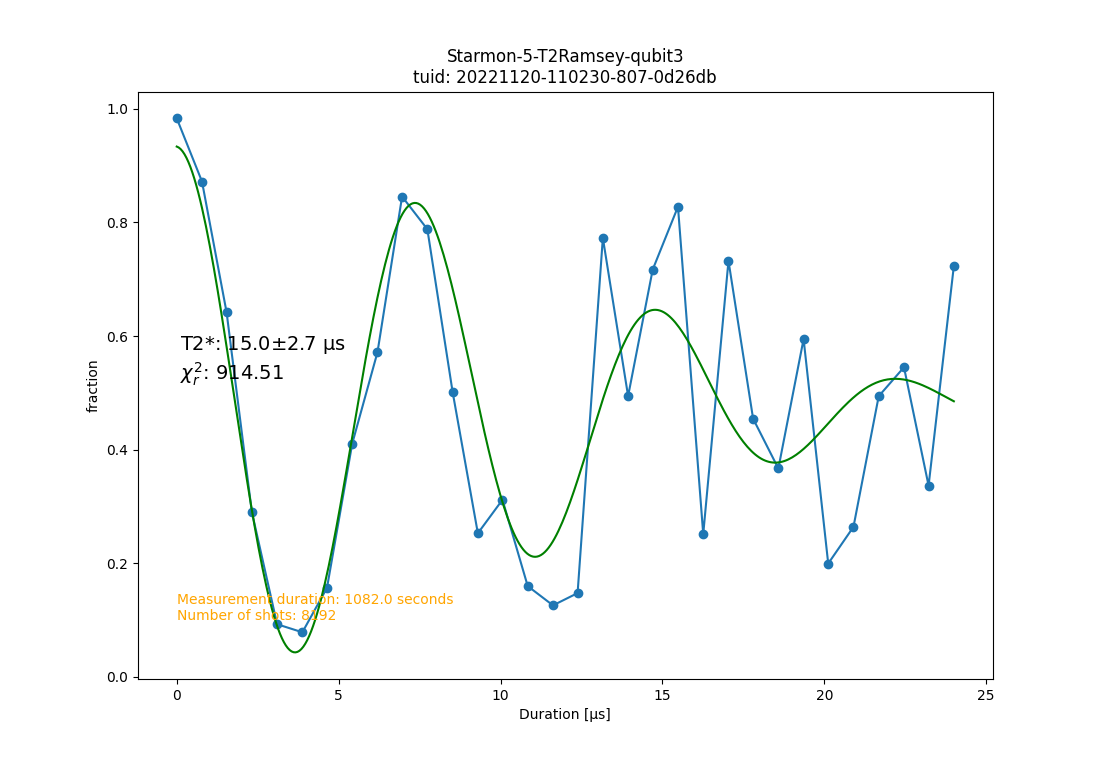}
  \caption{$T_2^*$ measurement on Starmon-5. The data points for small waiting time are good, for large waiting time the data seems non-coherent. The estimated value for $T_2^*$ itself is very reasonable, but
  the model violation $\chi^2_r$ is very high. Blue: measured data points, green: fitted model. }
  \label{figure:starmon5-q3-glitch-n2}
\end{figure}

The second example is a series of $T_2^*$ measurements on the Starmon-5 device
of Quantum Inspire~\cite{Last2020}. Over a period of a few weeks measurements of the $T_2^*$ coherence time have been  measured for all the 5 qubits.
A boxplot with the data is in figure~\ref{figure:t2star_boxplot}. There is variation
of the $T_2^*$ between the qubits, but also for a single qubit there is quite some variation of the
value with some strong outliers.
In figure~\ref{figure:chi2_vs_t2star} the model violation has been plotted for the different qubits
versus the time.
Qubit number 3 and 4 show poor fits.
We picked one of the outlier points with $\chi^2_r=914$ and $T_2^*=15.0$ us. The data for this measurement is shown in figure~\ref{figure:starmon5-q3-glitch-n2}.

\section{Discussion}

The \OLS performs as a robust all-round fitting method.
For the other methods such as \WLS and \MLE there are implementation choices that have to be made carefully, or the methods will result in a higher bias or outliers. Using correctly regularized and bootstrapped \WLS and \MLE can result in better variance of the estimate, but the difference is small for a higher number of data points or a high number of shots.

The model violation score is a useful tool for adding a quality estimate to fitted data.

\section{Acknowledgements}

This work was conducted with financial support from the Dutch National Growth Fund, which is part of the European Union’s NextGenerationEU recovery fund. This work was supported by the European Union’s Horizon 2020 research and innovation programme under the Grant Agreement No.951852 (QLSI project).


\bibliographystyle{plain}
\bibliography{references.bib}

\clearpage
\appendix

\afterpage{\clearpage}

\section{Results for fitting an exponential model}
\label{section:exponential_model}

\begin{figure}[ht]
  \centering
    \includegraphics[width=0.7\textwidth]{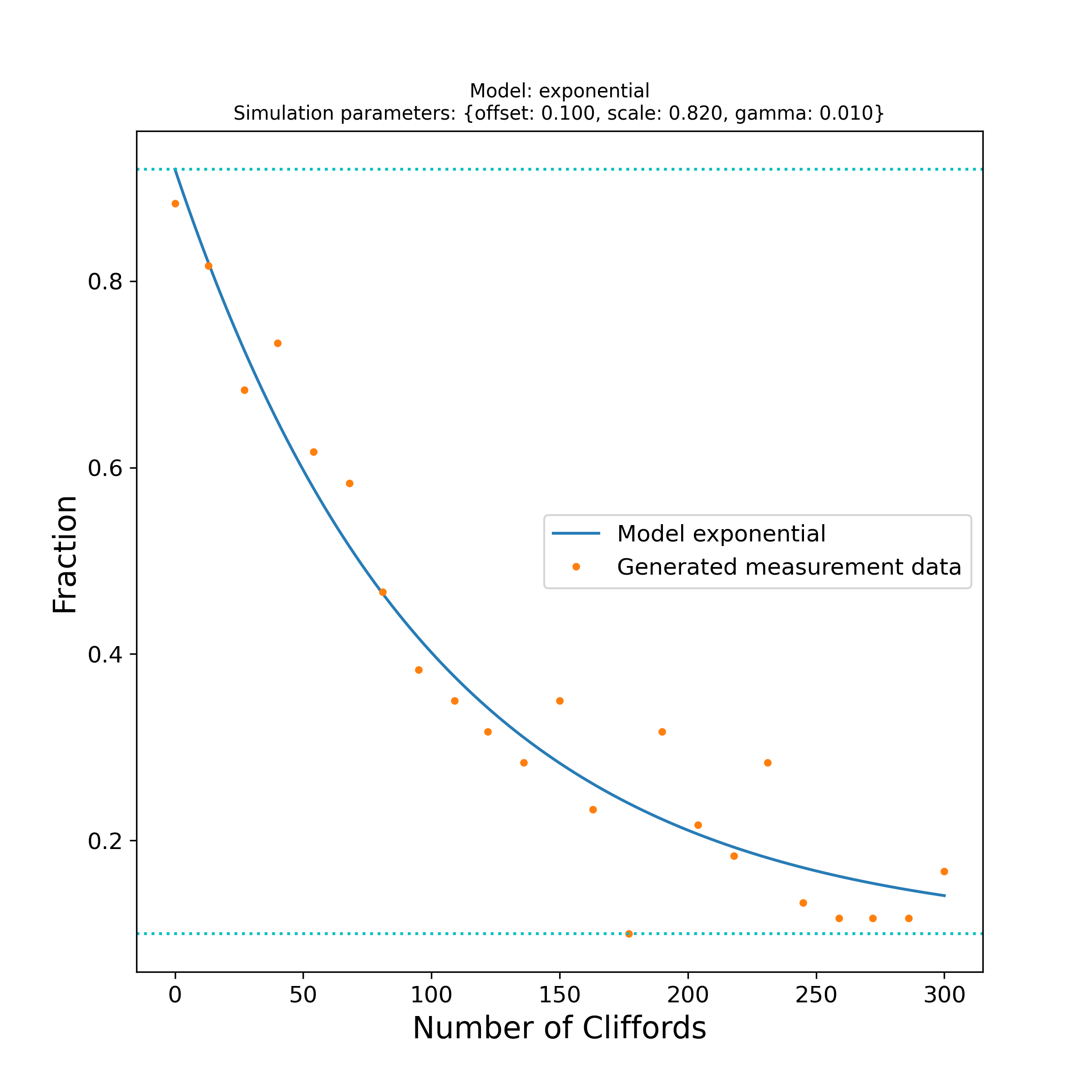}
  \caption{Model for exponential with simulated data}
  \label{figure:exponential model}
\end{figure}

\begin{figure}[ht]
  \centering
    \includegraphics[width=0.8\textwidth]{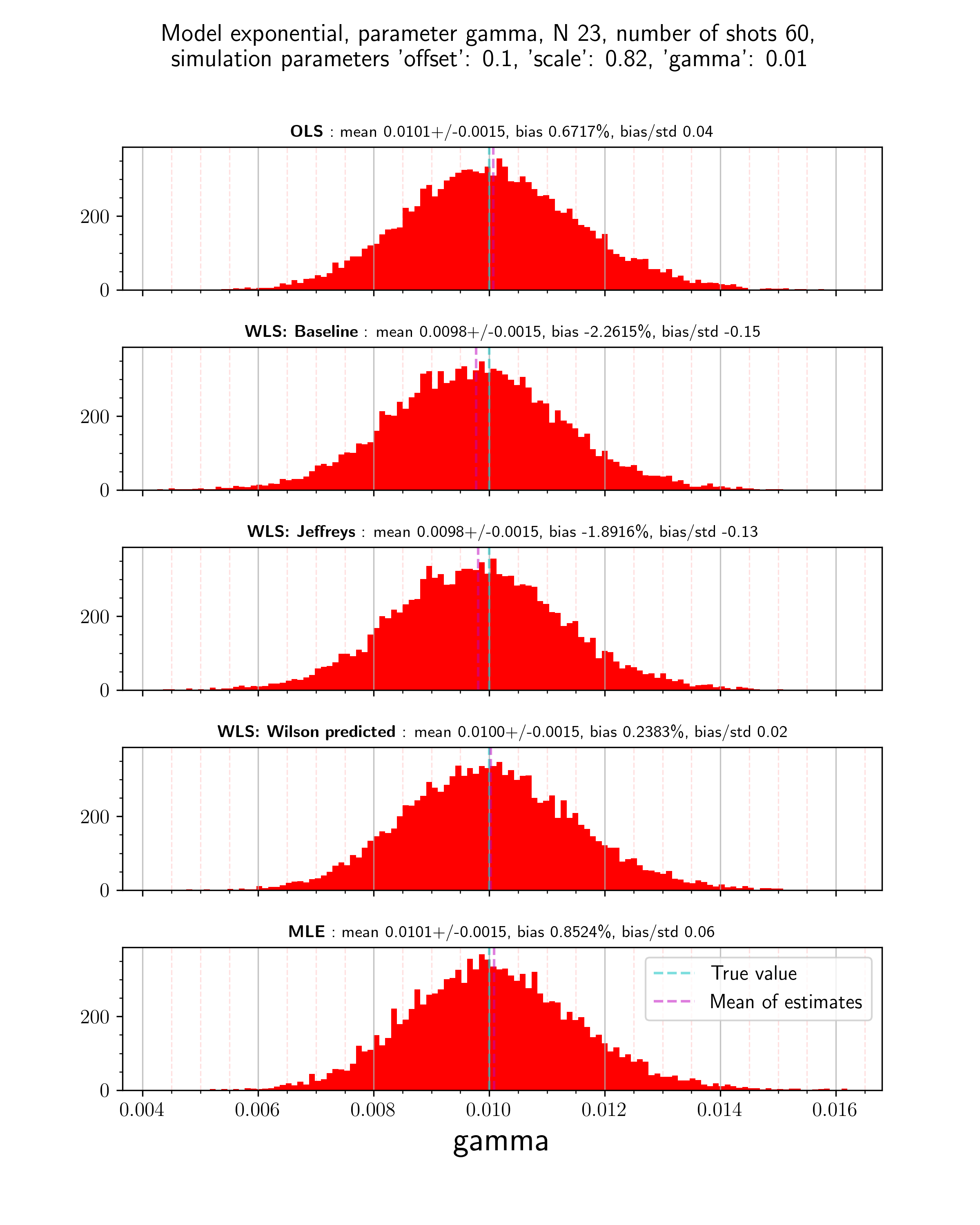}
  \caption{Distribution of the \code{gamma} parameter for an exponential model}
  \label{figure:fitted_parameter_distribution_exponential_gamma}
\end{figure}

\begin{figure}[ht]
  \centering
    \includegraphics[width=0.8\textwidth]{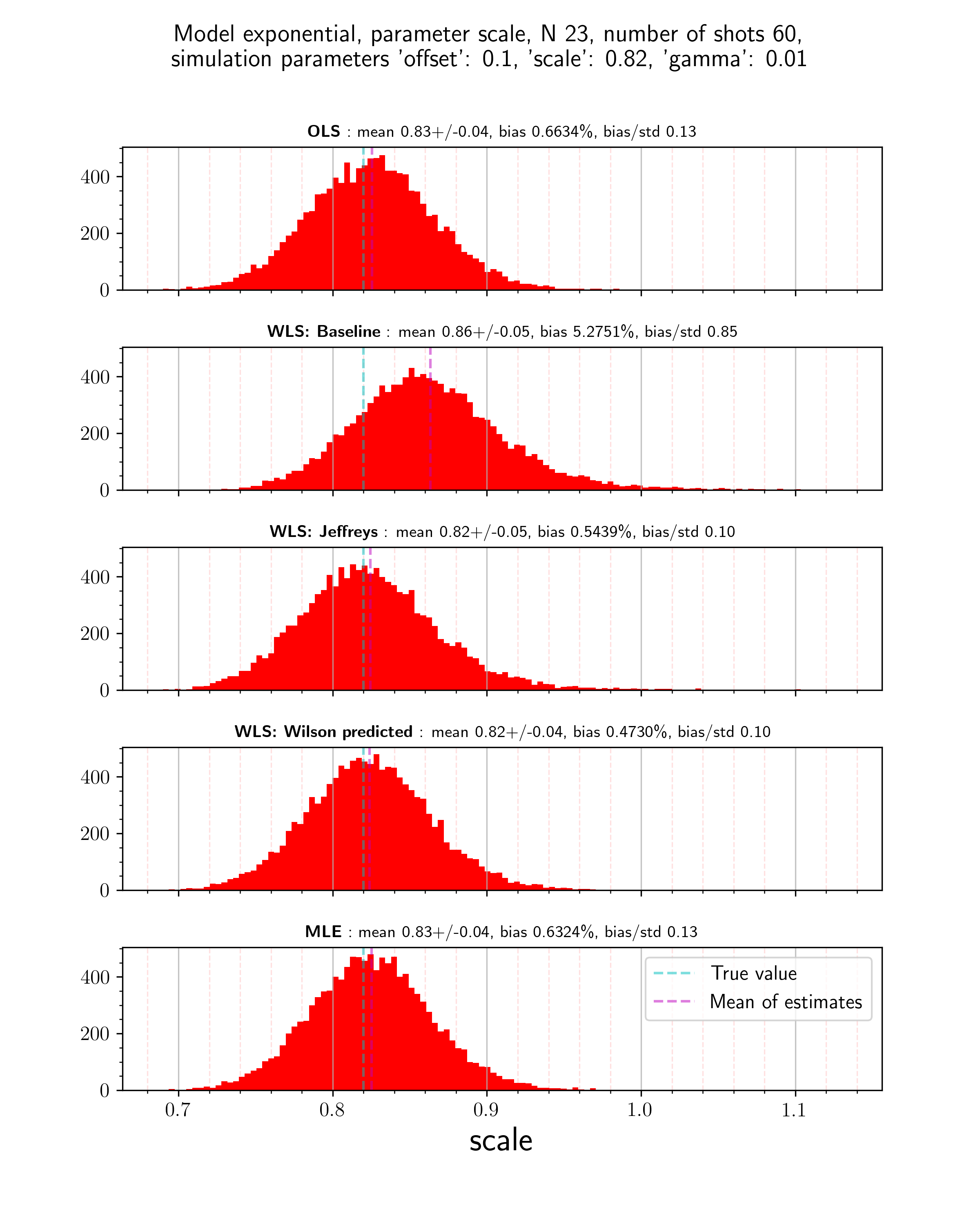}
  \caption{Distribution of the \code{scale} parameter for an exponential model}
  \label{figure:fitted_parameter_distribution_exponential_scale}
\end{figure}

\begin{figure}[ht]
	\centering
	\includegraphics[width=0.8\textwidth]{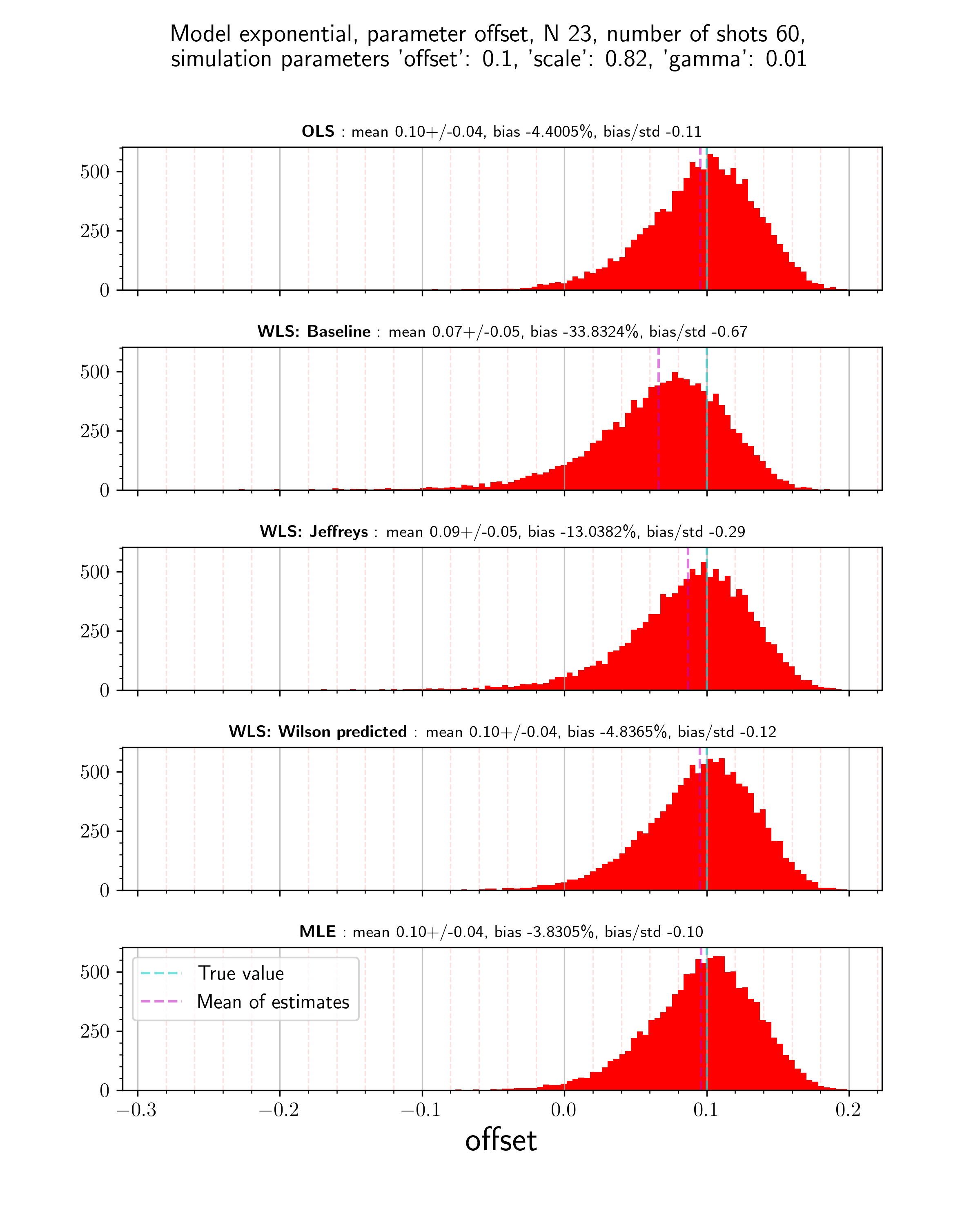}
	\caption{Distribution of the \code{offset} parameter for an exponential model}
	\label{figure:fitted_parameter_distribution_exponential_offset}
\end{figure}

In this section we show  results for fitting an exponential model $F(x, p) =p_0 + p_1 \exp(-p_2 x)$ with parameters $p=( \mathrm{offset} , \mathrm{scale}, \gamma)$. The model and generated data points are shown in~\figureref{figure:exponential model}.
All methods are identical to the methods for the results of the {sine} model
described in section~\ref{section:results}.
In \figureref{figure:fitted_parameter_distribution_exponential_gamma},
\figureref{figure:fitted_parameter_distribution_exponential_scale} and
\figureref{figure:fitted_parameter_distribution_exponential_offset} the distributions for
the $\gamma$, scale and offset parameters are shown.
The results for the various fitting methods are comparable to the sine fitting results.

\subsection{Maximum-likelihood fitting without penalty term}
\label{section:mle_penalty}

\begin{figure}[ht]
  \centering
    \includegraphics[width=0.8\textwidth]{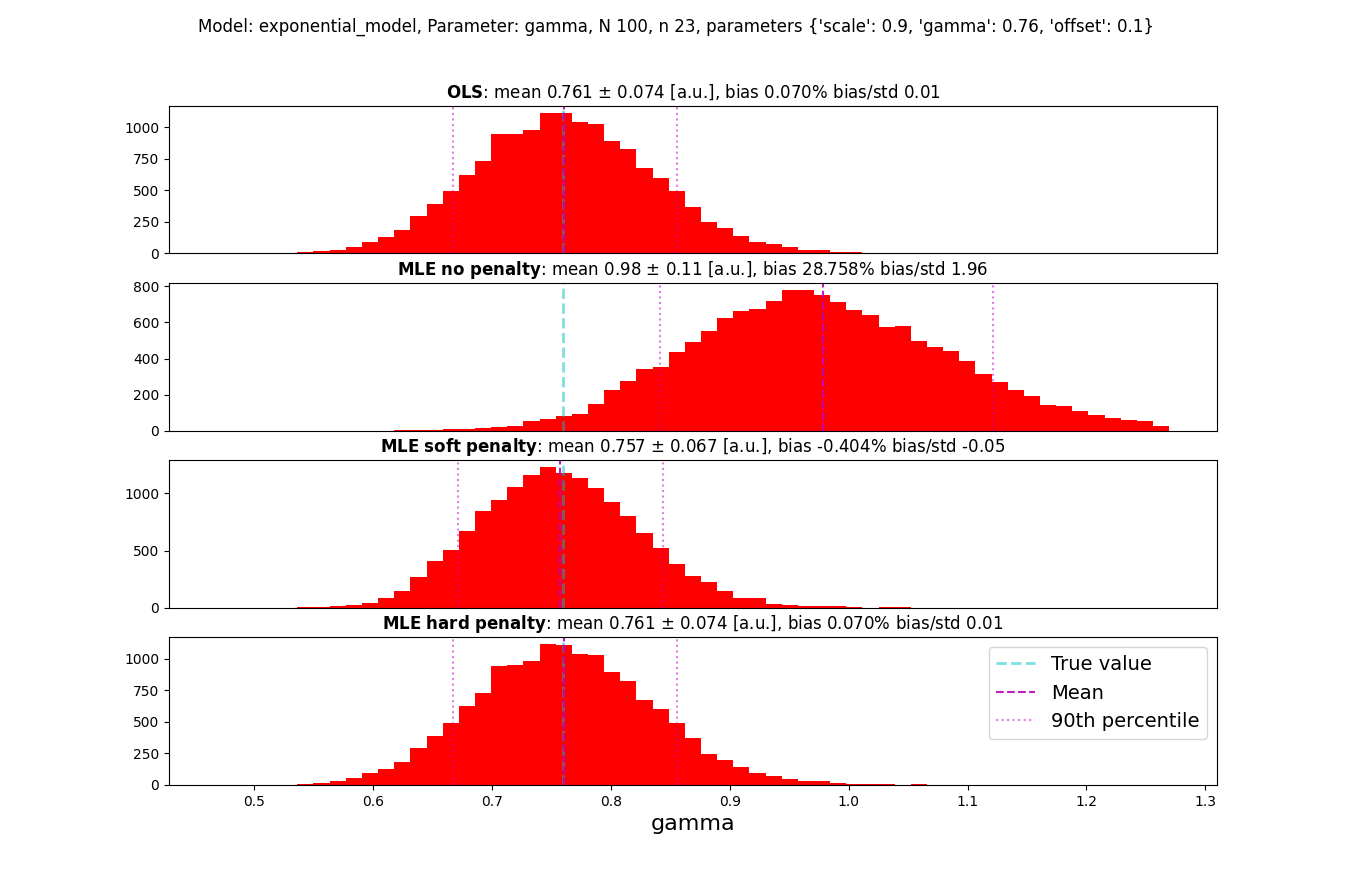}
  \caption{Distribution of the \code{gamma} parameter for an exponential model}
  \label{figure:mle_penalty}
\end{figure}

For the exponential model we compare the \OLS and \MLE with various penalty terms.
The distribution of the $\gamma$ parameter is in figure~\ref{figure:mle_penalty}.
It is clear that without a penalty term there is a strong bias in the result. The
soft and hard penalty have comparable results.

\section{Variance estimation details}

\begin{figure}
	\centering
	\includegraphics[width=0.8\linewidth]{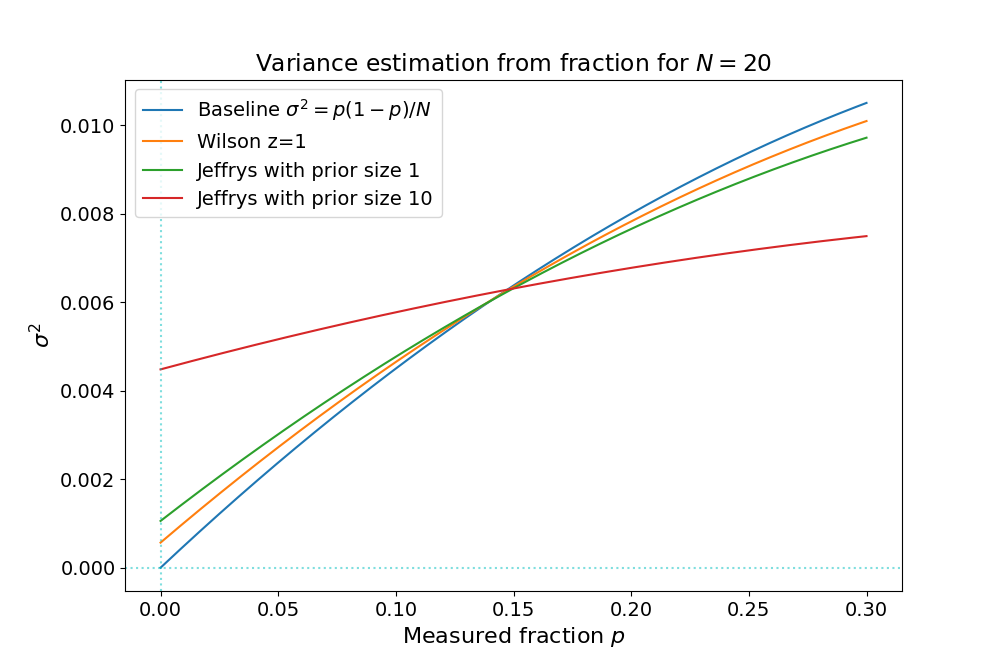}
	\caption{Different estimators for the variance from a measured fraction}
	\label{fig:varianceestimatefromfraction}
\end{figure}

From a measured fraction we can estimate a variance in the datapoint. Different options are shown in~\figureref{fig:varianceestimatefromfraction}.

\begin{qidoc}
	
	\pagebreak
	\section{Implementation in QLSI}
	\label{section:qlsi_implementation}

	The results of this paper are available in the \code{qi-analysis} package.
	The least-squares and weighted least-squares fitting is available
	in \code{qi\_analysis.fitter.fitter} (and various subclasses for specific models).
	The \code{qi\_analysis.fitter.mle\_fitter.MLEFitter} and \code{qi\_analysis.fitter.quantum\_wls\_fitter.QuantumWLSfitter} can be used for MLE and WLS fitting, respectively.
	To perform model testing the \code{qi\_analysis.statistics.model\_violation.ModelViolation} is available.
	
	The base fitting class is the \href{https://gitlab.com/qutech-sd/quantum-inspire/qi-analysis/-/blob/dev/src/qi_analysis/fitter/fitter.py#L120}{\code{LmFitter}} class.
	This class has various subclasses to fit specific models. Examples
	are the \code{SineFitter} or \code{ExponentialFitter}.
	An example of the usage to fit a simple model to data is:
	\begin{pythoncode}
		import numpy as np
		from qi_analysis.fitter import add_fitting_box, SineFitter
		from qi_analysis.data import load_qi_dataset 
		
		dataset = load_qi_dataset('6qubit_4l_cz_phase_crosstalk')
		fitter = SineFitter()
		result = fitter.fit_dataset(dataset)
		fitter.plot_fit(dataset, result)
		add_fitting_box(result)
		print(result)
	\end{pythoncode}
	The result of the code above is the output below, together with the image in figure~\ref{figure:qlsi_sine_fit}.
	\begin{pythoncode}
		LmFitResult(
		description='qi_analysis.fitter.sine.SineFitter',
		dataset_tuid='',
		fitted_parameters={
			'amplitude': 0.46434168392271347,
			'frequency': 0.15886752665476273,
			'phase': 5.5027031340176125,
			'offset': 0.4850618363486544
		},
		reduced_chi_squared=0.0003579070411619551
		)
	\end{pythoncode}
	
	\begin{figure}[ht]
		\centering
		\includegraphics[width=0.7\textwidth]{pictures/qlsi_sine_fit.png}
		\caption{\color{blue} Default output of the QLSI fitter class. The input data is shown together with the fitted model. The parameters shown are the parameters of the model together with the reduced $\chi^2$ value of the fit.}
		\label{figure:qlsi_sine_fit}
	\end{figure}
	Example Jupyter notebook showing how the fitting code works can be found
	at \href{https://qutech-sd.gitlab.io/quantum-inspire/qi-analysis/notebooks.html}{https://qutech-sd.gitlab.io/quantum-inspire/qi-analysis/notebooks.html}.

	\section{Additional results}

\end{qidoc}

\end{document}